\documentclass[prl, twocolumn,english,showpacs,floatfix]{revtex4}

\usepackage{graphics}
\usepackage{graphicx}
\usepackage{epsfig}
\usepackage{amssymb}
\usepackage{amsmath}
\usepackage{color}
\begin{document}

\newcommand{\cred}{\color{red}}
\newcommand{\cblue}{\color{blue}}
\title{Spectroscopy of ultracold atoms by periodic lattice modulations}

\author{C. Kollath$^{1}$, A. Iucci$^{1}$, T. Giamarchi$^{1}$, W. Hofstetter$^{2}$, U. Schollw\"{o}ck$^{3}$}
\affiliation{$^1$ Universit\'{e} de Gen\`{e}ve, 24 Quai
Ernest-Ansermet, CH-1211 Gen\`{e}ve, Switzerland}
\affiliation{$^2$ Institut f\"ur Theoretische Physik, J. W. Goethe-Universit\"at Frankfurt, 60438 Frankfurt am Main, Germany}
\affiliation{$^3$ Institute for Theoretical Physics, RWTH Aachen, D-52056 Aachen, Germany}

\begin{abstract}
We present a non-perturbative analysis of a new experimental
technique for probing ultracold bosons in an optical lattice by
periodic lattice depth modulations. This is done using the
time-dependent density-matrix renormalization group method. We find
that sharp energy absorption peaks are not unique to the Mott
insulating phase at commensurate filling, but also exist for
superfluids at incommensurate filling. For strong interactions the
peak structure provides an experimental measure of the interaction
strength. Moreover, the peak height of the peaks at $\hbar \omega \gtrsim 2U$ can be
employed as a measure of the incommensurability of the system.
\end{abstract}

\maketitle

The experimental realization of Bose-Einstein condensation in weakly
interacting ultracold atoms opened the way to numerous exciting
experiments directly probing fundamental effects of quantum
mechanics. More recently, the regime of strongly interacting atoms
has become experimentally accessible by Feshbach resonances
\cite{InouyeKetterle1998} and the advent of ultracold atoms in
optical lattices \cite{OrzelKasevich2001,GreinerBloch2002}. In a
pioneering experiment, the quantum phase transition from a bosonic
Mott insulator to a superfluid has been demonstrated
\cite{GreinerBloch2002}. More generally, ultracold atoms in optical
lattices open up the possibility to simulate and explore complex
quantum many-body phenomena known from electronic solids, like
high-temperature superconductivity, in a new context
\cite{hofstetter02}. One major advantage of ultracold atom systems
compared to their condensed matter counterparts is the tunability of
the system parameters. In particular, fast tunability in time made a
whole area of new non-equilibrium phenomena experimentally
accessible
\cite{GreinerBloch2002,GreinerBloch2002b,StoeferleEsslinger2004}.
The theoretical description of these phenomena, especially beyond
linear response, is still lacking. To probe quantum states in cold
atoms, various new measurement techniques have been developed.
Light-induced Bragg scattering yields the dynamical structure factor
\cite{StengerKetterle1999}, while RF absorption provides information
about single-particle excitations \cite{ChinGrimm2004}.
Time-of-flight expansion gives access to the momentum distribution
and correlations via detection of the average density and noise
\cite{AltmanLukin2003}.
Nevertheless, there is still a lack of measurement techniques
compared to condensed matter setups.

A qualitatively new way of probing the system which exploits fast
tunability was introduced  by St\"oferle \emph{et al.\ }
\cite{StoeferleEsslinger2004}. They determined
the excitation spectrum of ultracold bosons in an optical lattice by
measuring the heating induced by a relatively strong periodical
modulation of the lattice height (20\% of the initial lattice
height). Three distinct peaks at different frequencies were observed
in the energy absorption. Up to now, these features, in particular
the second and the third peak, are not well understood. Previous
theoretical studies applied a linear response treatment
\cite{IucciGiamarchi2005,ReischlUhrig2005,
PupilloBatrouni2006,BatrouniDenteneer2006} in analogy to condensed
matter systems or considered weak interactions using the
Gross-Pitaevskii equation \cite{KraemerDalfovo2005}. However, the
relatively strong modulation and the presence of a trapping
potential in the experiments, which implies inhomogeneous filling
\cite{JakschZoller1998,KashurnikovSvistunov2002}, demand for alternative
methods to check the validity of previous approximations.

In this Letter we present what is to our knowledge the first
simulation of the experiment by St\"oferle \emph{et al.} taking the
full time-dependence into account for reasonable system sizes in the
regime of strong and intermediate interaction. Hereby, we focus on
one spatial dimension using the adaptive time-dependent
density-matrix renormalization group method (adaptive t-DMRG)
\cite{DaleyVidal2004,WhiteFeiguin2004}. It is a numerical method
that allows for real-time evolution of quantum many-body systems out
of equilibrium with an explicitly time-dependent Hamiltonian. For
the case of commensurate filling we compare for weak modulations our results to the analytical
results of Iucci et al. \cite{IucciGiamarchi2005} to show the reliability
of our method. For large modulations saturation effects occur and reduce the height of the absorption
peaks. We show that at low temperature incommensurate regions with
more than one particle per site are crucial to reproduce
the multiple peak structure of the experimental absorption spectrum.
The appearance of the peak structure is not a specific sign of the
Mott insulating state as widely believed, but reflects the gap
occurring for strong repulsion between the coupled energy bands
(Hubbard bands). For strong interactions the peak positions can be
used as an experimental measure of the interaction strength. We also
show that the height of the peak at frequencies $\hbar \omega
\gtrsim 2 U$ can be employed as a measure of the incommensurability
of the system. Such a measure is highly relevant for the realization
of quantum computing with ultracold atoms in optical lattices, since
for current proposals the formation of wide regions with
commensurate filling is crucial.

The Bose-Hubbard model \cite{FisherFisher1989,JakschZoller1998}
\begin{equation*}
 H= -J \sum_{j=1}^{L-1} \left(b_j^\dagger b^{\phantom{\dagger}}_{j+1}+h.c.\right)
 + \frac{U}{2} \sum _{j=1}^{L} \hat{n}_j ( \hat{n}_j-1)+ \sum_{j=1}^{L} \varepsilon_j \hat{n}_j,
\end{equation*}
describes well ultracold bosons in a one-dimensional system subjected to an
optical lattice.
Here $L$ is the number of sites in the chain, $b^\dagger_j$ and
$b_j$ are the creation and annihilation operators, and $ \hat{n}_j=
b^\dagger_j b^{\phantom{\dagger}}_j$ is the number operator on site
$j$. The parameters $J$ and $U$ are the hopping amplitude and the
onsite interaction strength. The third term models the chemical
potential or an external potential, like the trapping potential. In
a system with commensurate filling a quantum phase transition occurs
at a finite ratio $(U/J)_c$ between a Mott insulating phase, in
which the atoms are strongly localized at the lattice sites, and a
superfluid phase with delocalized atoms. For incommensurate filling,
however, it is energetically unfavorable to localize all the atoms
and the system remains superfluid. The parameters of the
Bose-Hubbard model are directly related to experimental quantities.
For large lattice sizes an approximate formula is given by
\cite{Zwerger2003} $J/E_r= (4/\sqrt{\pi}) (V_x/E_r)^{(3/4)} \exp{(-2
\sqrt{V_x/E_r })}$ and $U/E_r= 4 \sqrt{2\pi}(a_s/\lambda) (V_x
V_\perp^2 /E_r^3)^{(1/4)}$. Here $E_r$ is the recoil energy, $a_s$
is the $s$-wave scattering length, and $\lambda$ is the wavelength
of the laser of the optical lattice. $V_x$ denotes the height of the
optical lattice and $V_\perp$ the strongly confined transverse
direction \footnote{Here we use the tight transverse confinement
$V_{\perp} \equiv 30E_r$.}. For a better comparison to the
experiment \cite{StoeferleEsslinger2004} we use
$a_s=5.45\textrm{nm}$  of $^{87}$Rb in the $F=2$, $m_F=2$ hyperfine
state and $\lambda= 825 \textrm{nm}$. A periodical modulation of the
lattice height, here we use $V_{x}(t)=V_{0}[1+\delta V \cos (\omega t)]$, translates into a periodic variation of the hopping
coefficient $J$ and the interaction coefficient $U$ via
$J(t)=J[V_x(t)]$ and $U(t)=U[V_x(t)]$. 
\begin{figure} [ht]
  \begin{center}
 \includegraphics[width=0.7\linewidth]{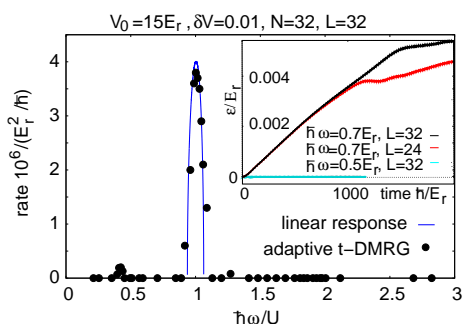}
  \end{center}
  \caption{    Absorption rate vs.\ the modulation frequency $\omega$ at a lattice depth $V_0=15 E_r$; the corresponding values of the Bose-Hubbard model are $U\approx 0.71E_r$ and $J\approx 0.0074E_r$, i.e. $U/J\approx 95$.
    The lines are analytical results obtained by
    linear response treating the hopping term as a perturbation
  \cite{IucciGiamarchi2005}. The symbols are the results obtained using the
  adaptive t-DMRG. Inset: Energy absorbed versus time. $\omega =0.71E_r/\hbar$
     corresponds to a modulation frequency at resonance and $\omega
    =0.5E_r/\hbar$ away of resonance. }
  \label{fig:complin}
\end{figure}

In Fig.~\ref{fig:complin} the dependence of the energy absorption
rate on the modulation frequency is shown for a system which is
initially in the Mott insulating phase with filling $n=1$ per
lattice site \footnote{
In our adaptive t-DMRG calculation the explicit time-dependence of  the Hamiltonian
was approximated by a sequence of small time steps with time- independent Hamiltonians. Step sizes (down to
$0.2 \hbar/E_r$) were determined by a convergence analysis. We keep  several hundred states in the effective
Hilbert space (see Ref.~\cite{GobertSchuetz2004} for a detailed error analysis).}.
We found this rate $\alpha$ to be independent of the system size before
saturation occurs (see inset Fig.~\ref{fig:complin}). It is
determined by fitting the function $f(t)= E(t=0)+\alpha t+ b_1
\cos(\omega t+b_2)$ with the fitting parameters $\alpha$, $b_1$, and
$b_2$ to our results before saturation. The rate
shows a two peak structure: a large peak at $\hbar \omega \approx U $
and an approximately 20 times smaller one at $\hbar \omega \approx
U/2$. 
The peak at $\hbar \omega
\approx U$, in the following called $U$ peak, corresponds to
particle-hole excitations by a ``single photon'' process
\cite{StoeferleEsslinger2004,IucciGiamarchi2005,ReischlUhrig2005,PupilloBatrouni2006}.
For a small modulation amplitude $\delta V = 0.01$ the results for
the $U$ peak obtained by linear response combined with perturbation
theory in $J/U$ (solid lines) agree very well with our full
calculation. 
The peak at $\hbar \omega \approx
U/2$, which does not occur in linear response, is due to
``two-photon'' processes where twice the energy quantum is needed to
generate a particle-hole excitation \cite{Foot2005}. 
In principle
absorption peaks at frequencies which are higher multiples of $U$
can also arise. For commensurate filling, our calculations
demonstrate that they are negligible \footnote{This can also be
understood from a perturbative calculation: to first order only the $U$ peak has non-vanishing weight in a
commensurately filled system; the linear response matrix elements at
$2U$ vanish as $(J/U)^3$. The $3U$ peak results from
two contributions: i) three particle-hole excitations, and ii) two
particle hole excitations with both particles jumping on the same
site. The matrix element of the process i) vanishes faster than
$J/U$ and therefore only the process ii) contributes to the $3U$ peak at
this order.}.

Up to now we considered an initially Mott insulating state. However,
if the filling is incommensurate, the system remains superfluid for
all interaction strengths. In the limit of strong interaction,
energy bands (Hubbard bands) occur separated by an energy
gap of order $U$, but the system remains superfluid, since
excitations in the highest occupied band are gapless. 
Therefore, as pointed out before
\cite{KraemerDalfovo2005,IucciGiamarchi2005,PupilloBatrouni2006},
incommensurability changes the low ($\omega \ll U/\hbar$) frequency spectrum. However, the presence of a trap and the
experimental resolution makes it difficult to observe this change in
experiment. We show how incommensurability also imprints clear
signatures in the high ($\omega \sim 2U/\hbar$) frequency spectrum.

\begin{figure} [ht]
  \begin{center}
     \includegraphics[width=0.65\linewidth]{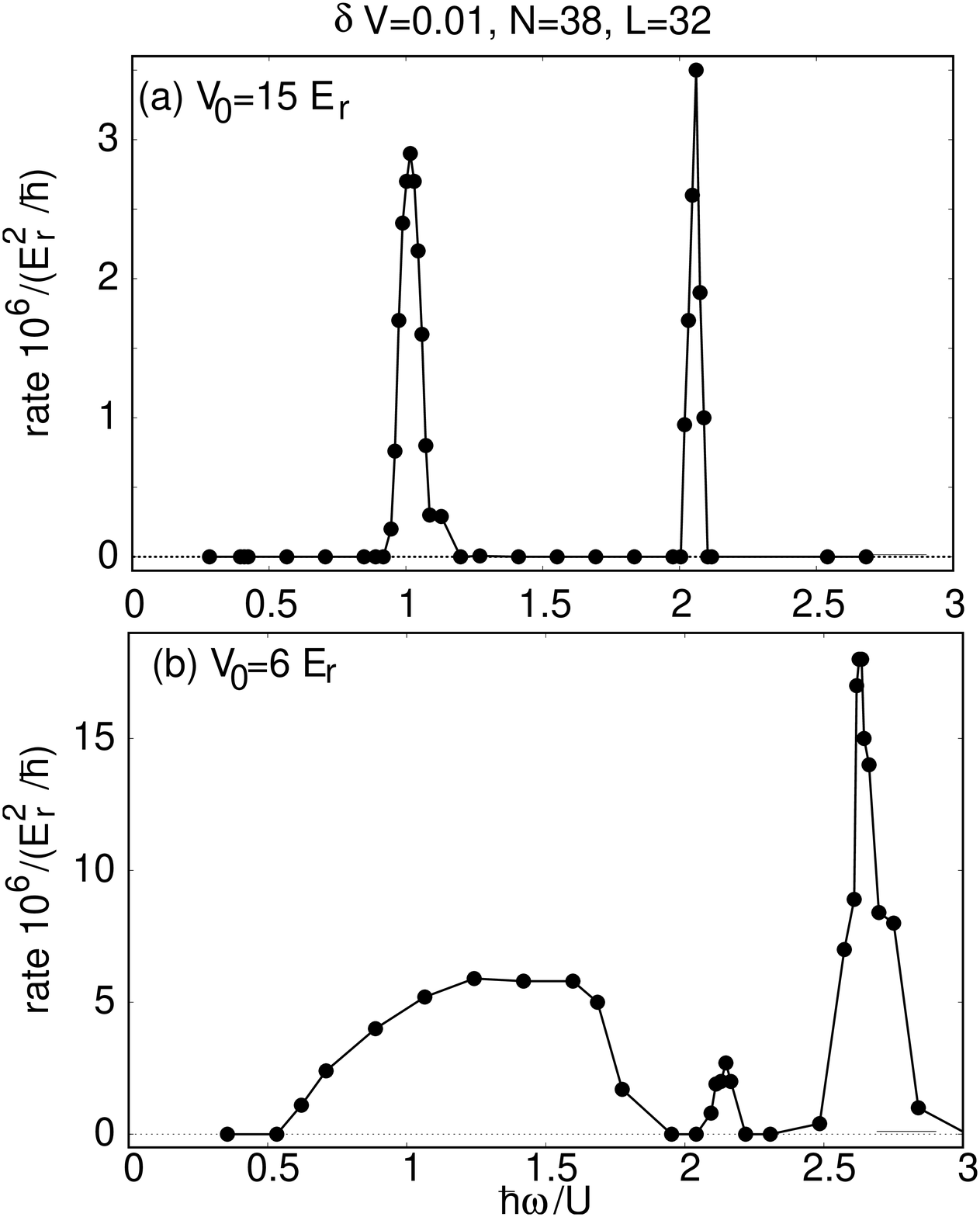}
   \end{center}
  \caption{Energy absorption rate vs.\ modulation frequency $\omega$ at incommensurate filling $N=38$
  particles on $L=32$ sites for a
    lattice height of (a) $V_0=15 E_r$ and (b) $V_0=6E_r$. Compared to the
    system at commensurate filling additional peaks at $\hbar \omega \approx
    2U$ and in (b) at $\hbar \omega \approx
    2.6 U$ arise. Solid lines are guide to the eyes.
   }
  \label{fig:incom}
\end{figure}

For a deep lattice, see Fig.~\ref{fig:incom} (a), in addition to the
$U$ peak a new resonance at $\hbar \omega \approx 2U$, in the
following called $2U$ peak, arises \footnote{The peak at $\hbar
\omega \approx U/2$ is much smaller and cannot be seen here. The
height of the peak at $\hbar \omega \approx U$ decreases compared to
the commensurate case (cf. Fig.~\ref{fig:incom} and
\ref{fig:complin}).}. The main contribution to it stems from the
process in which one particle hops onto an already doubly occupied
site. The process where two particle-hole excitations are created is
negligible. Therefore the $2U$ peak is a clear indicator for regions
of the optical lattice with incommensurate filling. For the
homogeneous system considered here we find that even at zero
temperature the $2U$ peak for an incommensurate filling $n\approx
1.2$ is much higher than the $U$ peak. Finite temperature could
generate further defects in the occupation and thereby additional
contributions to the $2U$ peak as noted in \cite{ReischlUhrig2005}
for the commensurate system \footnote{In a commensurate system a
high temperature of $T=U/3$ is necessary to generate a second peak
of considerable height \cite{ReischlUhrig2005}. This temperature is
much higher than the one expected in the experimental setup.}. The
peak height of the $2U$ peak could be taken as a measure of the
`degree' of incommensurability present in the system regardless of
the origin of the incommensurability.
The same finding holds for larger values of the hopping parameter.
The peak structure becomes more complicated [cf.
Fig.~\ref{fig:incom} (b)], but the peaks at $\hbar \omega \approx 2U$ and $\hbar \omega \approx 2.6
U$ still only appear for incommensurably filled systems. The shift
in position compared to the naively expected positions at integer
multiples of $U$ can be understood from the change in the energy
structure. Whereas for strong interactions narrow energy bands with
a gap of $U$ exist, for intermediate interactions the bands broaden
with increasing hopping. In particular, the ground state of the
incommensurate system shifts down strongly with increasing hopping
strength. As a result, the energy difference towards excited states
increases which causes in Fig.~\ref{fig:incom} (b) the shift of the center of the $U$ peak to approximately
$1.2U$ and of the $2U$ peak to approximately $2.6U$. The peak around $2.1U$
stems from a splitting of the $U$ band in processes in which a particle
hole pair is created on singly occupied sites and doubly occupied sites. The ratio of the amplitudes of the peaks depends crucially on
the degree of incommensurability in the system. We have verified these findings by exact diagonalization of Bose-Hubbard chains up to 7 sites.

\begin{figure} [ht]
  \begin{center}
    \includegraphics[width=0.7\linewidth]{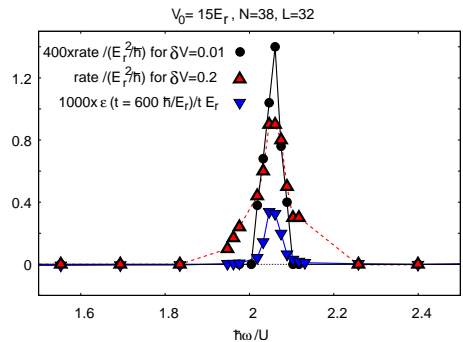}
      \end{center}
  \caption{Saturation effects for strong modulations. The absorption rate for a weak modulation (a) $\delta
    V_1=0.01$ (circles) and a strong modulation (b) $\delta V_2=0.2$ (upward
    triangles) are compared. The rate for (a) is scaled by the
    ratio $(\delta V_2/\delta V_1)^2$ to remove trivial scaling. Clear
    saturation effects in the peak
    height and width can be seen.
Saturation of the integrated energy absorption in time:
(c) is the integrated energy absorbed up to time
    $t_m=600 \hbar/ E_r$ per unit time (downward triangles) , i.e. $ [E(t_m)-E(0)] /t_m$ for $\delta V_2=0.2$. The chosen
    value of $\delta V_2=0.2$ and $t_m=600\hbar/E_r$ correspond to the
    experimental values. Due to saturation effects in time it is
    smaller than the energy absorption rate shown in (a).
 }
  \label{fig:compdV02}
\end{figure}

In order to use this $2U$ peak as a measure of the amount of
incommensurability, one needs to determine it for the strong
modulations ($\sim 20\%$) used in the existing experiments
\cite{StoeferleEsslinger2004}. The adaptive
t-DMRG method, which has no problem dealing with the strong time
dependent modulation, is thus ideally suited to tackle this question
whereas it is delicate to use linear response for this purpose. To
illustrate the difference we compare in Fig.~\ref{fig:compdV02} the
energy absorption rate for the case of a small modulation $\delta
V_1=0.01$ and a modulation with the strength $\delta V_2=0.2$.
The results for $\delta V_1$ are rescaled by a factor of $(\delta
V_2 / \delta V_1)^2$ to eliminate the trivial amplitude dependence
expected from linear response. Although the structure of center of the peak
agree qualitatively well even for these high modulation strength,
saturation effects occur in particular in the height of the peaks. 
Our results indicate that linear response overestimates the actual
energy absorption. For the parameters shown here a reduction by a
factor of approximately $1.6$ takes place. A bimodal structure seems
  to appear for strong modulations which causes a broadening of the peaks,
  here by a factor of approximately 2. We also plotted
the experimentally measured quantity, namely the heating, i.e. the
integrated absorbed energy. In Fig.~\ref{fig:compdV02} we show
$[E(t_m)-E(0)]/t_m$ for a modulation time $t_m=600 \hbar /E_r$
chosen as in the experiment. The positions of the peaks agree well
in the two spectra but the height deviates, for the parameter shown
here approximately by a factor of $2$. This is mainly due to saturation effects in time (cf.~inset Fig.~\ref{fig:complin}) which cause the actual integrated
absorption per unit time to remain lower than the rate \footnote{In contrast the
width of the absorption peaks is smaller than the width found for
the rate with the same modulation strength ($20\%$) and resembles
the width found for the rate of a smaller modulation strength
($1\%$). A detailed study of the saturation effects goes beyond
the scope of the work presented here.}.

\begin{figure} [ht]
  \begin{center}
    \includegraphics[width=0.75\linewidth]{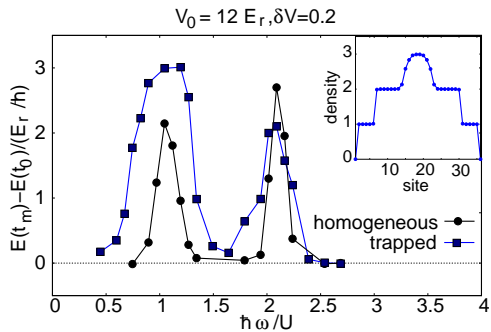}
      \end{center}
  \caption{The integrated energy absorbed up to time
    $t_m=50 \hbar/ E_r$, i.e. $ [E(t_m)-E(0)]$ for $\delta
 V_2=0.2$ in a system with confining potential and N=65 (squares) and in a
 homogeneous system with N=46 and L=32 (circles). The inset shows the initial
 density distribution in the trapped system. 
 }
  \label{fig:trap}
\end{figure}
In order to relate our results to present day experiments we have to
 investigate how much the trap affects the energy absorption spectrum. We find that
 the positions of the peaks are robust against the presence of a trapping
 potential \footnote{This is expected by the simple picture in which particle hole
  excitations are created on neighbouring sites. There the trapping only
  causes a shift of the excitation energy by the difference of the site
  on which the particle hole pair is generated. Since the sign of the shift
  depends on which site the particle and on which the hole is created it will
  lead to a simple broadening.} and that the amplitudes remain of the same order of magnitude. 
This findings can be seen in Fig. \ref{fig:trap} in which we show an example of the integrated absorbed energy at a fixed time
$t_m$ versus the modulation frequency for the case of the presence of an
incommensurate region in the center of the parabolic trap (see inset
 Fig.~\ref{fig:trap})\footnote{We used a potential of the form $V_t=0.006(j-L/2-1/2)^2$. The results for the initial state may have large
  uncertainties at the boundaries of locked phases, since here many
  states almost degenerate in energy exist. Nevertheless, this should not
  change the qualitative
  picture found. Due to computational reasons the time $t_m$ is chosen
smaller than in the experiment. We found that by this we underestimate the
amplitude of the $2U$ peak compared to the $U$ peak compared to the experiment, since it grows with time $t_m$. }.
Therefore as in the homogeneous case the occurence of a peak at $\hbar
\omega \approx 2U$ signals the presence of an incommensurate region and our findings enables us to
compare the positions in the homogeneous system (Fig.~\ref{fig:incom}) to the experimental data
\cite{StoeferleEsslinger2004}.

In the experiment for large values of
the initial lattice height narrow peaks at $\hbar
\omega \approx U$ and $\hbar \omega \approx 2U$ were found. These
peaks broaden and shift in energy if the lattice becomes more
shallow. For intermediate interaction strength an additional peak at
$\hbar \omega \approx 2.6 U$ appeared. These findings for the
positions of the peaks agree excellently with our results at zero
temperature \emph{provided} we assume the presence of an
incommensurately filled region in the experimental system (cf. Fig.~\ref{fig:incom}).

To conclude we have demonstrated using the adaptive t-DMRG method
that the measurement procedure \cite{StoeferleEsslinger2004} gives
important information about the properties of the bosonic system, in
particular about commensurability properties and the energy levels.
It will therefore be of great interest to extend these calculations
and measurements to cold fermions in optical lattices, which have
only recently been realized experimentally
\cite{KoehlEsslinger2005}.

We would like to thank M. Cazalilla, A. Ho, S. Huber, K.
Schmidt, and C. Tozzo for fruitful discussions and the group of T. Esslinger for
stimulating discussions and for sharing their experimental data. This
work was partly supported by the Swiss National Fund under MaNEP and
Division II, and by the DFG grant HO 2407/2-1.


\end{document}